\documentclass[a4paper,cits]{PoS}
\usepackage{graphics, graphicx}
\usepackage{amsmath}
\usepackage{amssymb}
\usepackage[authoryear,square]{natbib}
\bibpunct{(}{)}{;}{a}{}{,}
\title{Understanding pulsar magnetospheres with the SKA}

\ShortTitle{Pulsar magnetospheres}
\author{\speaker{A. Karastergiou$^1$},
{S. Johnston$^2$},
{N. Andersson$^3$},
{R. Breton$^4$},
{P. Brook$^1$},
{C. Gwinn$^5$},
{N. Lewandowska$^6$},
{E. Keane$^7$},
{M. Kramer$^8$},
{J.-P. Macquart$^9$},
{M. Serylak$^{10}$},
{R. Shannon$^2$},
{B. Stappers$^{11}$},
{J. van Leeuwen$^{12}$},
{J. P. W. Verbiest$^{13}$},
{P. Weltevrede$^{11}$},
{G. Wright$^{14,11}$}
\\
$^1$ Oxford Astrophysics, Denys Wilkinson Building, Keble Road, Oxford, OX1 
 3RH, UK.\\  E-mail: \email{aris@astro.ox.ac.uk}\\
$^2$ CSIRO Astronomy \& Space Science, Australia Telescope National Facility, PO Box 76, Epping, NSW, Australia\\
$^3$ Mathematical Sciences and STAG Research Centre, University of Southampton, Southampton SO17 1BJ, UK\\
$^4$ School of Physics and Astronomy, University of Southampton, Southampton SO17 1BJ, UK.\\
$^5$ Department of Physics, University of California, Santa Barbara, CA 93106, USA\\
$^6$ Astronomy Department, University of W\"urzburg, Emil-Fischer Strasse 31, 97074 Würzburg, Germany\\
$^7$ Centre for Astrophysics and Supercomputing, Swinburne University of Technology, Mail H30, PO Box 218, Hawthorn, VIC 3122, Australia\\
$^8$ Max-Planck-Institut fuer Radioastronomie, Auf dem Huegel 69, 53121 Bonn, Germany.\\
$^9$ ICRAR/Curtin Institute of Radio Astronomy, Bentley, WA 6845, Australia\\
$^{10}$ Department of Physics \& Astronomy, University of the Western Cape, Private Bag X17, Bellville 7535, South Africa\\
$^{11}$ Jodrell Bank Centre for Astrophysics, The University of Manchester, Alan Turing Building, Manchester, M13 9PL, United Kingdom.\\
$^{12}$ ASTRON, the Netherlands Institute for Radio Astronomy, Postbus 2, 7990 AA Dwingeloo, The Netherlands\\
$^{13}$ Fakultaet fuer Physik, Universitaet Bielefeld, Postfach 100131, D-33501 Bielefeld, Germany\\
$^{14}$ Astronomy Centre, University of Sussex, Falmer, Brighton BN1 9QJ\\
}

\abstract{ The SKA will discover tens of thousands of pulsars and
  provide unprecedented data quality on these, as well as the
  currently known population, due to its unrivalled sensitivity. In
  this chapter, we outline the state of the art of our understanding
  of magnetospheric radio emission from pulsars and how we will use
  the SKA to solve the open problems in pulsar magnetospheric
  physics. Pulsars are known to be variable on numerous timescales:
  from nanoseconds to decades, corresponding to events, likely
  associated with the emission process itself through to discrete
  stable magnetospheric states. Timescales in between show phenomena
  such as pulse-to-pulse variations, sub-pulse drifting and nulling in
  operation, in addition to the relevant orbital, cooling and magnetic
  field decay timescales. The increased sensitivity and frequency
  range of the SKA will allow us to study these effects with higher
  fidelity than ever before. Moreover, it will be possible to perform
  these studies down to the individual pulse level for a large sample
  of millisecond pulsars. With their significantly different magnetic
  field strength and magnetospheric extent, this will provide
  important clues to the magnetospheric physics. Monitoring the SKA
  pulsar population, by fully exploiting its sub-arraying and
  multi-beaming capabilities, will enable us to study all of these
  phenomena and gain quantitative information on pulsar
  magnetospheres. Polarization profiles of a large number of pulsars,
  and their evolution over a broad frequency range, will enable us to
  perform measurements, with unprecedented detail, of the geometry of
  the emission regions of these systems. In addition to studying the
  plethora of new sources, we will perform intense targeted studies of
  already known objects. For example, with the SKA we will get a
  unique view of the magnetospheres of the two stars in the Double
  Pulsar System, down to the precision of a single rotation period,
  over a wide range of frequencies with full polarization
  information. We will be able to map the magnetospheres of selected
  sources by performing high sensitivity observations from 50 MHz all
  the way through to gamma-rays. Furthermore, using scintillation
  imaging techniques, we will be able to resolve pulsar magnetospheric
  features. With the good sample of radio-emitting magnetars, which
  the SKA pulsar surveys will provide, we will be able to compare, and
  identify any magnetospheric differences, between these and the
  standard pulsar population. As well as finally settling the
  questions about the pulsar emission mechanism and magnetospheric
  structure once and for all, we can use this understanding to improve
  the usefulness of pulsars as astrophysical tools. For example,
  understanding the switching behaviour in intermittent pulsars that
  is likely to be in operation in all pulsars at some level, and the
  detailed pulse-to-pulse modulation properties, will enable improved
  calibration of pulsar signals for use as precision clocks, which
  aides in the SKA's headline science quests involving precision
  pulsar timing.}

\FullConference{
Advancing Astrophysics with the Square Kilometre Array\\
June 8-13, 2014\\
Giardini Naxos, Italy}

\newcommand{\degr}[1]{\ensuremath{#1^\circ}}

\newcommand{\skipthis}[1]{}
\begin{document}

\section{Pulsar magnetospheres and the SKA}

Pulsar radio emission is used as a high precision tool in a number of
astrophysical experiments. Currently, the most high profile
experiments involve tests of gravity through pulsar timing, and the
related search for a stochastic gravitational wave background. To take
full advantage of the intrinsic rotational stability that
characterizes pulsars, it is necessary to model all the effects that
perturbe the measured pulse times-of-arrival. Over the past decade,
the importance of modelling radio pulse propagation through the
interstellar medium has been understood, and improvements are now
becoming evident in timing residuals. This has revealed the next level
of ``timing noise'', most likely related to processes that occur near
and within the radio emitting magnetosphere. For the first time, it is
becoming clear that the lack of understanding of the pulsar radio
emission mechanism is presenting obstacles to experiments that use
pulsars as tools.

The quest to improve this situation with the SKA, relies on its
ability to deliver the following elements:

\noindent {\bf 1.} high sensitivity, to observe individual pulses from
pulsars at a high signal-to-noise ratio;\\
{\bf 2.} broadband receivers and backends, to study the frequency
dependence of emission phenomena;\\
{\bf 3.} polarization, to better understand the pulsar geometries, the
structure of the pulsar radio beams and the details of emission and
propagation in the pulsar magnetosphere.\\
{\bf 4.} high-cadence monitoring, by observing multiple sources in
sub-array mode, to uncover the physics that govern variability on all
timescales.

The sole motivation to understand the radio emission mechanism is not
to improve timing experiments. A steady stream of science output in
areas such as relativistic plasmas in the magnetosphere, variability
due to internal or external triggers, pulsar winds and supernova
physics has resulted from the study of details of the radio emission
mechanism on numerous interesting pulsars. The SKA surveys will
deliver a new population of radio pulsars, among which we expect a
group of sources that are ideal to address specific aspects of the
emission mechanism problem (e.g. pulsars with interpulses for
polarimetric studies of pulsar geometry).

In this chapter, we present the state of the art in observations
aiming at understanding the magnetospheres of pulsars.  We address
this topic from three angles:

\noindent {\bf 1.} Understanding the three dimensional (3D) structure of the radio emission beam;\\
{\bf 2.}Understanding how magnetospheres evolve between different classes
of pulsars;\\
{\bf 3.} Unlocking the information in the characteristic observational
timescales of pulsar emission.

\section{The radio emission beam}
Our current understanding of the emission mechanism which operates in
the magnetospheres of radio pulsars is limited. Progress can be
achieved by determining where in the magnetosphere the radio emission
is generated. Understanding where charged particles are accelerated to
relativistic energies, would provide a better picture of the global
pulsar electrodynamics, an essential element in solving the puzzle of
how pulsars operate. In this section it is described how radio
observations with the SKA are expected to advance our understanding of
the elusive radio emission mechanism by allowing the construction of a
3D map of the pulsar magnetosphere.

\subsection{Pulsar geometry and polarization}
\label{sectionPulsarGeometryAndPolarization}
Radio polarization measurements allow geometrical properties of the
radio beam to be quantified. Basic viewing geometry parameters include
the magnetic inclination angle (angle between the magnetic dipole axis
and the rotation axis of the neutron star) and the angle between our
line-of-sight and the rotation axis. Since the early days of pulsar
astronomy, the way the position angle of the linear polarization 
changes during the rotation of the star (PA-swing) is interpreted
with the rotating vector model \citep[][RVM]{rc69a,kom70}, which
directly depends on these two angles. Another important parameter of
the emission beam, which can be derived using polarization measurements,
is the radio emission height: co-rotation of the emitting region
causes the PA-swing to be delayed with respect to the pulse profile,
the magnitude of the shift being determined by the emission
height~\citep{bcw91}.

Currently, the viewing geometries have been derived for about a
hundred objects \citep[see e.g.][]{ran90,gou94}, while emission
heights are determined for only a fraction of those
\citep[e.g.][]{mr01,ran93,wj08}. There are a number of complications
which limit the precision and the fraction of the total population of
pulsars suitable for this kind of analysis. Firstly, emission is only
observed for a small interval of the rotation of the star. This limit
makes a large range of RVM solutions become degenerate and in
virtually all cases the viewing geometry cannot be determined from RVM
fitting alone. In order to get unique solutions, additional
assumptions are necessary. However, this introduces large and poorly
understood systematics. Drawing statistical conclusions about the
population as a whole is, therefore, known to be problematic, and
emission heights determined in different ways often do not agree
\citep[see e.g.][]{ml04,wj08}.  Secondly, many pulsars do not have
RVM-like PA-swings, showing that our understanding of pulsar
polarization is incomplete. This severely limits the sample suitable
for studying their viewing geometry. Moreover, because young pulsars
are more suitable for this kind of analysis \citep[see
e.g.][]{jw06,wj08} and references therein), the sample is necessarily
biased.  Thirdly, the above method to determine emission heights
requires knowledge about which point in the radio profile corresponds
to the closest approach of the line of light to the magnetic
axis. Therefore knowledge about how the emission regions are
distributed within the beam is crucial.

There is currently only one pulsar for which RVM fitting alone provide
us with a unique viewing geometry without making additional
assumptions~\citep{kj08}. This was possible because this pulsar has an
interpulse (both magnetic poles are observed, which provides more
rotational phase coverage) and it is young (therefore the PA-swing
obeys the RVM very precisely). Simply observing currently known
pulsars with a greater sensitivity is not expected to be effective, as
systematic errors are the problem. The key is to find more of these
``ideal case'' pulsars, which then can be used to find out which
additional assumptions are justified allowing emission geometries to
be determined for a larger sample.

The SKA will discover many additional pulsars, including a
sufficiently large ``clean'' sample of pulsars with RVM-like PA-swings
to allow statistical studies of the pulsar population. As described
above, this will be a biased sample in terms of their location in the
$P$-$\dot{P}$ diagram. By comparing the results of the ``clean''
sample with the results of a larger sample the poorly understood
biases can be investigated in detail. This ultimately will lead to a
better understanding in the physical processes which are currently
missing in models, thereby advancing our understanding of the emission
mechanism.

As shown for instance by~\cite{dyk08}, non-RVM-like PA-swings can
teach us about the deformations of the magnetic field structure due to
magnetospheric currents and rotational sweep-back of the magnetic
field lines \citep[see e.g.][]{dh04}. The determination of emission
geometries can also be used to determine the way the magnetic
inclination angle changes over time~\citep{tm98,wj08a,ycbb10}, which
provides information about the torque acting on the pulsar. In
addition, propagation effects of the radio waves in the pulsar
magnetosphere are expected \citep[e.g.][]{bp12}. Better understanding
will allow the relevant plasma parameters of the emission mechanism to
be determined. Good determination of emission geometry is also important
for the study of drifting subpulses, as it
allows the emission fluctuations observed for our line-of-sight to be
related to a changing beam structure over time \citep[see e.g.][]{dr01,vj97}.

\subsection{Radio emission regions}
\label{secEmissionRegions}
Despite the limited success at building a theoretical framework for
the interpretation of pulsar radio emission, empirical models that
attempt classifications of emission characteristics have uncovered
observational patterns that any theory should be able to account
for. For example, in a series of papers by Rankin and collaborators
\citep{ran83,ran90,ran93}, evidence is presented that points to radio
emission emanating from sets of active magnetic field lines, forming
one, two or more hollow cones of emission. Different heights then
correspond to different frequencies, in accordance with the
observation that pulse profiles often become narrower towards high
frequencies \citep[e.g.][]{tho91a}. In addition to the hollow cones,
Rankin's classification included a type of component originating close
to the magnetic axis which was labelled core. These often show
distinguishing emission properties such as steeper spectra and
complicated polarization features.

In contrast, \cite{lm88} showed that the two-dimensional organization
of radio components of a large population of
pulsars is patchy, and although their characteristics broadly match
the Rankin classification, there is no strong evidence of two emission
mechanisms for core and cone emission. It was also noted that some
young pulsars have observable emission from both magnetic poles. This
property, coupled with the wide, relatively simple profiles of young
pulsars and their polarization characteristics, was used by \cite{jw06}
to suggest that radio emission originates from high altitudes in young
pulsars compared to the older pulsar population. This idea was
expanded further \cite{kj07} to produce a phenomenological model that
statistically explains the complexity of profiles from young and old
pulsars (not millisecond pulsars). The main element of this model
includes emission from a small number of discrete multiple heights for
older pulsars, following ideas presented in \cite{gg01}. Multiple
emission heights have implications on geometrical modeling.

Broadband observations, polarization and sensitive single pulse
observations are all essential to understanding the detailed 3D
location of radio emission regions in pulsar magnetospheres.

\subsection{Radio spectra and broadband nature of emission}
The radio emission of pulsars is produced by relativistic particles in
the circumpulsar plasma. Pulsar spectra must be explained by a model
which describes the emission mechanism, in combination with the energy
distribution of the particles responsible for the observed emission
\citep[e.g.][]{mm80}. Therefore, detailed measurements of pulsar
spectra provide valuable information about how the radio emission is
produced.  In addition, understanding the pulsar spectrum in a
statistical sense for the pulsar population is crucial for population
synthesis to refine models for the parent distribution of pulsars,
which gives rise to the observed distribution~\citep{blv12}.

The spectra of radio pulsars are broadband and resemble (at least to
first order) power laws with a relatively steep frequency
dependence. Above $\sim$200 MHz the average spectral index was shown
by \cite{mkk+00} to be $-1.8$, with a standard deviation of 0.2
(i.e. pulsars are brighter at lower frequencies), although
\cite{blv12} suggest a mean spectral index closer to $-1.4$ with unity
standard deviation. The dependence on pulsar parameters, such as $P$
and $\dot{P}$, is weak, but correlations with characteristic age are
found~\citep{lylg95}. Understanding the way spectra depend on the spin
parameters will give insight into how magnetospheric conditions change
throughout the $P-\dot{P}$ diagram.

In general, deviations are observed from this first order behavior, one
of which is that at lower frequencies, the spectrum often ``turns
over'' \citep[e.g.][and references therein]{sie02}.
It is currently unknown if this is
because of a decrease in efficiency of the emission mechanism, or
because of an absorption mechanism becoming effective. It is also not
known why some pulsars do not have these low-frequency turnovers, but
others do. There is a hint that these spectral turnovers do not happen
for millisecond pulsars~\citep{kl01}, suggesting that their
magnetospheric conditions are significantly different compared to
normal (non-recycled) pulsars. Apart from the lack of a spectral break
at low frequencies the spectra of MSPs and slow pulsars are remarkably
similar~\citep{kxl+98}. At higher frequencies ($\sim10$ GHz or
above) spectral breaks are observed, such that the spectrum flattens
or even turns up~\citep{wjkg93,kjdw97}.

A complication in the interpretation of radio spectra is that the
observed spectrum is affected by a geometric feature. This feature
arises because the beam shape is frequency dependent such that we see
different parts of the beam at different frequencies.  An
understanding of this effect is therefore crucial
to infer the intrinsic spectra from those observed.

\subsection{The interstellar medium as an interferometer of resolved
  pulsar magnetospheres}
Interstellar scintillation affords the potential to resolve the tiny
emission regions of pulsars.  By virtue of the compactness of the
emission, pulsar radiation is subject to strong interference effects
as it propagates through the turbulent plasma of the Interstellar
Medium (ISM). However, pulsar magnetospheres have a finite size and,
for extended sources, each ray path produces a subimage.  The
radiation observed at Earth is the result of the interference between
many thousands of sub-images, or speckles, across the scattering
region.  Since the scale of the speckle pattern exceeds $\sim 10$\,AU
in some cases \citep[e.g][]{bmg+10}, interstellar scattering offers
the prospect of achieving {\emph picoarcsecond} angular resolution.
Scintillation-based tests of the emission size fall into three broad
categories according to the manner in which the finite size of the
emission region manifests itself on the radiation properties.

An angular displacement in the emission site causes a lateral
displacement in the scintillation pattern at Earth.  If that
displacement is parallel to the direction of motion of the pulsar, and
the turbulence is ``frozen'', then a lag in time produces precisely
the same effect on the scintillation pattern as a change in pulse
phase.  This technique was applied by \cite{wc87} during a double
imaging event to place an upper limit on the emission size of
PSR\,B1237+25 of $10^6$~km, roughly an order of magnitude larger than
its light cylinder.

Although the different lines of sight from the source are coherent,
the different parts of the subimages are not.  For a point source,
interference forms a diffraction pattern, a ``scintillation pattern'',
in the observer plane.  The intensity varies randomly from zero to
many times the average; indeed, zero intensity is the most common
value.  A resolved source forms a superposition of diffraction
patterns from each individual point on the source; these points are
not coherent, so that the observed pattern is the incoherent
superposition of the patterns from a point source.  The effect upon
the diffraction pattern depends on the size of the source, with a
typical scale of the angular resolution of the scattering region,
viewed as a lens, as the size of a source that affects the
distribution of intensity significantly.  The observed distribution of
intensity, or of interferometric visibility, provides a measure of the
size of the source.  Comparison of the pattern between two widely
spaced observatories can provide information on the shape of the
source.  The technique was used by \cite{gjr+12} to measure a size for
the emission region of the Vela pulsar of $<100$\ km to $~1000$\ km,
with the smallest size near pulse center and the largest at the
beginning and end.  A single-dish technique was used by \cite{jgd12}
to set an upper limit of $<40$\ km, averaged over the entire pulse.
The difference in measured size may reflect the emergence of an
additional, ``conal'' component at the shorter wavelength, the broader
pulse, or other factors.

The technique of speckle scintellometry extends the technique that
deduces the emission region structure based on the lateral
displacement of the scintillation pattern by using holographic
techniques to further boost the S/N of the pulsar signal by partially
descattering the radiation \citep{wkss08}.  Application of this
technique to PSR\,B0834+06 yields a determination of the astrometric
phase shift across the pulse profile equivalent to an angular
resolution of 150\,picoarcseconds, or 10\,km at the distance of the
pulsar \citep{pmdb14}.

\subsection{Neutron Star Precession}
\subsubsection{Mapping pulsar beams}
\begin{figure}
  \begin{center}
    \includegraphics[width=0.40\textwidth, angle=-90]{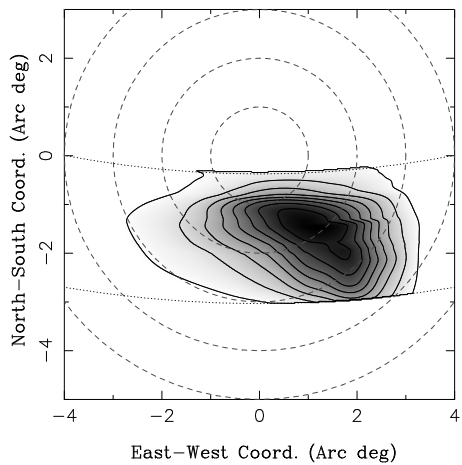}
\end{center}
\caption{\label{fig1141} Two-dimensional shape of the emission beam
  over the traversed region in PSR J1141--6545. The inclination of the
  magnetic axis is taken to be $\degr{160}$. The dotted lines show the
  path traversed by the line of sight; the lower one corresponds to
  the beginning of the data span and the upper one to the minimum
  impact parameter reached around MJD 54000. For more details see
  \cite{mks+10}.}
\end{figure}

The observed pulse profiles show a large variety of shapes produced by
a one-dimensional cut through the two-dimensional intensity
distribution given by the pulsar beam.  In General Relativity, the
proper reference frame of a freely falling object suffers a precession
with respect to a distant observer, called geodetic precession. In a
binary pulsar system, this geodetic precession leads to a relativistic
spin-orbit coupling, analogous to spin-orbit coupling in atomic
physics.  As a consequence, the pulsar spin axis precesses about the
total angular momentum vector, changing the relative orientation of
the pulsar towards Earth. In such a case, we should expect a change in
the radio emission received from the pulsar, as first proposed by
\cite{dr74} very soon after the discovery of PSR B1913+16.  Indeed, as
our line-of-sight moves through the emission beam, each profile
represents a slightly different cut through the beam structure.  By
adding measured profiles in the right order, it is possible to
reconstruct a real 2-D map of a pulsar beam. One also expects that the
polarization will change with time.  Currently, such an experiment has
been successfully applied to PSRs B1913+16 \citep{wt02,cw08,kra02},
J1141$-$6545 \citep{mks+10} (see Fig. \ref{fig1141}), J1906+0746
\citep{kas12} and recently J0737$-$3039B \citep{pmk+10}.  The results
are so far inconclusive as to whether a systematic pattern exists.

\subsubsection{Free precession}
Free precession is the most general motion executed by a rotating
solid body. Given that neutron stars have elastic crusts (and may
contain solid components in their core, according to some nuclear
physics models), they can sustain asymmetries which may lead to the
system precessing.  In order to illustrate the main issues, let us
consider the case of PSR B1828--11, which provides the strongest
evidence in favour of free precession so far \citep{sls00}. Possible
precession solutions for the data, which shows periodicity at both
around 500d and 1000d, were first discussed in \cite{ja01,le01}. More
recently, it has been argued that the observed periodicity does not in
fact represent free precession but is a result of state-switching in
the magnetosphere \citep{lhk+10}. However, as argued in \cite{jon12},
this does not necessarily mean that precession does not play a role in
the system.

If we, for the moment, assume that PSR B1828--11 is freely precessing,
then \cite{ja01} argue that the data requires the magnetic dipole to
be very nearly orthogonal to the star's deformation axis (associated
with the asymmetry that forces the precession).  This would lead to an
inferred wobble angle in excellent agreement with the value estimated
from the amplitude modulation.  The data for PSR B1828--11 is at
variance with the level of superfluid pinning required to explain the
large glitches seen in the Vela pulsar. It also does not allow for the
expected pinning of superfluid vortices to magnetic fluxtubes in the
star's superconducting core \citep{lin03}. At the moment we seem to
have three options: i) Our understanding of neutron stars
superfluidity/superconductivity is not quite right \citep{lin03}, ii)
various instabilities may intervene and affect the fluid motion, thus
altering the conclusions \citep{gaj08}, or iii) we are not seeing free
precession \citep{lhk+10}.

\section{Energetics of pulsar magnetospheres and radio emission}
Within the pulsar population, there are particular populations with
clear and to some extent interpretable emission characteristics
\citep{tkb+14}. The SKA pulsar surveys will allow for new
breakthroughs \citep{kbk+14} by providing significantly more examples
from each of the populations described in the following.

\subsection{Young Pulsars}
The youngest pulsars, which have the highest spin-down energy loss
rates, appear to have different emission properties to the more
middle-aged pulsars.  In particular, they possess extremely high
levels of linear polarization \citep{jw06,wj08}, not generally seen in
the older pulsars.  Furthermore, the polarization position angles
follow the RVM curve to a much greater degree than the old pulsars do;
this is particularly important in determining their geometry (see
section 2.1).  Curiously, it appears as if the orientation of their
linear polarization is either parallel to the magnetic field or
perpendicular to it \citep{jhv+05,ran07b}. This can be understood in
terms of plasma physics, which predicts orthogonal polarization modes
\citep{mth75}. Some evidence suggests that the younger pulsars have
`simpler' profiles to those of the middle-aged pulsars \citep{kj07}.

The majority of young pulsars are also high-energy emitters. Since the
launch of the {\it Fermi} satellite more than 200 pulsars are now known to
be $\gamma$-ray emitters \citep{aaa+13}. The relationship between the
$\gamma$-ray emission and the radio emission is a critical one to our
understanding of the magnetosphere. Much progress has been made in
this area in recent times, but results are still inconclusive
\citep[see e.g.][]{wrwj09,rw10,pghg12}. This class of pulsars is likely to
grow in importance as more high-energy pulsars are detected.

\subsection{Magnetars}
Magnetars are a class of neutron stars composed of Anomalous X-ray
Pulsars (AXPs) and Soft Gamma-ray Repeaters (SGRs). Currently there
are 21 magnetars detected with 5 more awaiting confirmation
\footnote{\small http://www.physics.mcgill.ca/$\sim$pulsar/magnetar/main.html}.
Originally, SGRs were discovered to have bursting emission in the hard
X-ray/soft gamma-ray range \citep{mgi+79}, while AXPs are steady,
X-ray emitting sources with rotational periods exceeding few
seconds \citep{fg81}. Subsequent observations of powerful bursts from
AXPs \citep{gkw02}, as well as the discovery of persistent X-ray
emission and slow rotational periods from SGRs \citep{kds+98} have
revealed the common nature of their emission properties
\citep{dt92a}. Magnetars were typically known to share the following
properties: very strong magnetic fields exceeding the quantum critical
value for electrons ($B \sim 4.4 \times 10^{14}$\,G); decay of such
fields is believed to create the observable high X-ray and gamma-ray
luminosities \citep{td95,td96a}, often visible in bursts;
spin periods of 5\,--\,12 s with rapid spin-down on timescales 
of $10^{3}-10^{5}$ years; and absence of radio emission.

\begin{figure*}
\centerline{
\includegraphics[scale=.4,angle=0]{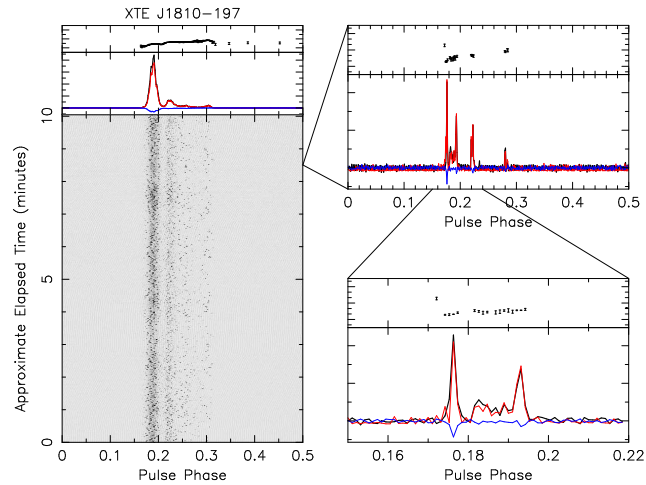}
}
\caption{{\bf Left}: {\bf Pulse stack from an} 8.35 GHz observation with
  the average profile and position angle swing plotted on top. Total
  intensity is plotted in black, while linear and circular
  polarization are plotted in red and blue respectively. {\bf Top
    right}: Individual single pulse profile with visible narrow and
  spiky emission.  {\bf Bottom right}: Close-up of the pulse plotted
  above. }
\label{fig_magnetar}
\end{figure*}

Recent discoveries of pulsars with spin periods and inferred dipolar
magnetic field strengths typically seen in magnetars
\citep{ckl+00,msk+03} and at which radio emission should not occur,
bring magnetars and pulsars closer as a single population. The
detection of transient radio emission from XTE J1810--197 and 1E
1547--5408 \citep{crh+06,crhr07}, followed by the discovery of PSR
J1622--4950, a magnetar first detected in radio \citep{lbb+10} with
subsequent identification of its X-ray counterpart \citep{ags+12},
further strengthens this link. Another example is SGR J1745--2900, the
only radio emitting magnetar found in the inner parsec of the Galactic
Centre \citep{efk+13,sj13}. The radio emission from magnetars is highly
variable in nature and usually declines in tandem with its higher
energy counterparts. Magnetar radio emission typically appears after
X-ray outbursts; has a flat spectral index \citep{ljk+08} which
enables detection even of individual pulses at frequencies above 88
GHz \citep{crp+07}; shows pulse profile morphology changing
dramatically on timescales of minutes to days
\citep{ccr+07,crj+07,lbb+12}; shows polarization similar to the
emission properties of normal radio pulsars but with differences which
can be explained as propagation effects in a non-dipolar magnetic
field \citep{crj+08,ksj+07}. Fig.~\ref{fig_magnetar} shows a typical
observation of XTE J1810--197 with the 100-m Effelsberg radio
telescope at 8.35 GHz.

The population of radio magnetars is very small at present. As they
emit over a wide spectrum of wavelengths ($\gamma$-ray, X-ray,
optical, near infrared and radio) studies at multiple frequencies are
best poised to provide information on these objects. The SKA pulsar
surveys will significantly increase the sample of radio emitting
magnetars and uncover their relationships with the other populations
of radio emitting neutron stars.

\subsection{Millisecond Pulsars}
Most of our current knowledge of the pulsar emission mechanism and
pulsar magnetospheres is derived from observations of slow
pulsars. Only a very limited number of the known millisecond pulsars
(MSPs) is bright enough to enable detailed studies of their
single-pulse and magnetospheric properties. However, the studies that
have been performed so far have shown interesting differences between
MSPs and slow pulsars, motivating detailed studies of MSPs at the high
sensitivity provided by the SKA.

The first clear difference between MSPs and slow pulsars is that the
RVM cannot explain the polarimetry of MSP emission, even though other
polarimetric properties (such as polarization degree and the presence
of orthogonally polarized modes), do not differ
\citep{ymv+11}. Similarly ambiguous results are found in the profile
variations as a function of observing frequency: even though the pulse
profiles of MSPs vary less strongly (which indicates a far smaller
emission region than in the case of slow pulsars), the overall changes
are reminiscent of slow pulsars, but shifted to higher frequencies
\citep{kll+99}.  Concerning intensity variations of single pulses,
\cite{jg04} found no variations (other than giant pulses) in PSR
B1937+21, while clear pulse-to-pulse modulations were found in PSR
J0437-4715 \citep{jg04,ovb+14}.  \cite{es03} discovered giant pulses
in several pulsars and possible sub-pulse drifting. For the brightest
MSP, PSR J0437-4715, the recent work by \cite{ovb+14} rules out
drifting sub-pulses.

In summary, a clear picture of the emission properties of MSPs has yet
to emerge. The lack in telescope sensitivity has limited this area of
research to essentially the two brightest MSPs, which are far from
representative of the whole MSP population in many regards. Early
indications do show that the emission mechanism for slow pulsars and
MSPs must be similar, if not identical. Understanding the emission
process in MSPs is an essential element in the continued research of
pulsar magnetospheres.

\subsection{The double pulsar}
The double pulsar PSR J0737$-$3039A/B is not only a remarkable system
in the context of testing gravity theories; it also provides a unique
window of opportunity to study pulsar magnetospheres. This binary
comprises a rapidly spinning 23-ms pulsar, {\em A}, in a 2.4-hr orbit
with a slowly rotating 2.8-s pulsar, {\em B}. The fortunate nearly
edge-on alignment of the orbit with our sight line is such that pulsar
A is eclipsed for $\sim20$~s when it passes behind pulsar B. The
duration implies a transverse cross-section of a few km at the
location of pulsar B from A, thus implying a magnetospheric origin. A
high time resolution study by \cite{mll+04} revealed that the eclipsed
signal from A is modulated by the rotation of B during the eclipse, a
phenomenon which led \cite{lt05} to interpret the eclipses as being
caused by synchrotron resonance in the closed field lines of pulsar
B. The modulation naturally arises from the change in optical depth as
the magnetic field geometry changes over the course of B's
rotation. Subsequently, \cite{bkk+08} demonstrated the success of this
model in a long-term study of the radio eclipses, which led to the
direct measurement of relativistic spin precession of pulsar B.

The double pulsar eclipses have two majors implication for the study
of pulsar magnetospheres. Firstly, they allow us to directly probe the
topology of a pulsar's magnetic field, which has been shown to be
primarily dipolar \citep{bkm+12}. Secondly, multi-frequency
observations of the eclipse profile can constrain plasma properties
such as the density, profile and multiplicity of the plasma. It has
been shown that the density profile drops abruptly at a radius well
inside the co-rotation radius (i.e. light cyclinder) and requires a
large multiplicity \citep{bkm+12}.  SKA1 will allow us to perform a
polarization study of the eclipses which will yield an independent
test of the eclipse mechanism and geometry of the system \citep{lt05}.
Polarization studies will also allow us to infer the strength of the
magnetic field and compare it to predictions from pulsar timing.

\section{Timescales in pulsar magnetospheres}
While it has been known since the early days of pulsar astronomy that
individual pulses are intrinsically variable, over the past decade a
number of important observational results have shown that both radio
emission and the magnetosphere itself can vary on timescales spanning
at least $18$ orders of magnitude, from nanoseconds to decades. In the
following, we describe the current understanding of radio emission
phenomenology on all timescales and how this relates to pulsar timing
experiments. It is important that SKA pulsar observation scheduling is
determined by our understanding of the timescales of magnetospheric
variability, in order to maximize scientific return.

\begin{figure*}
\centerline{
\includegraphics[scale=.5,angle=270]{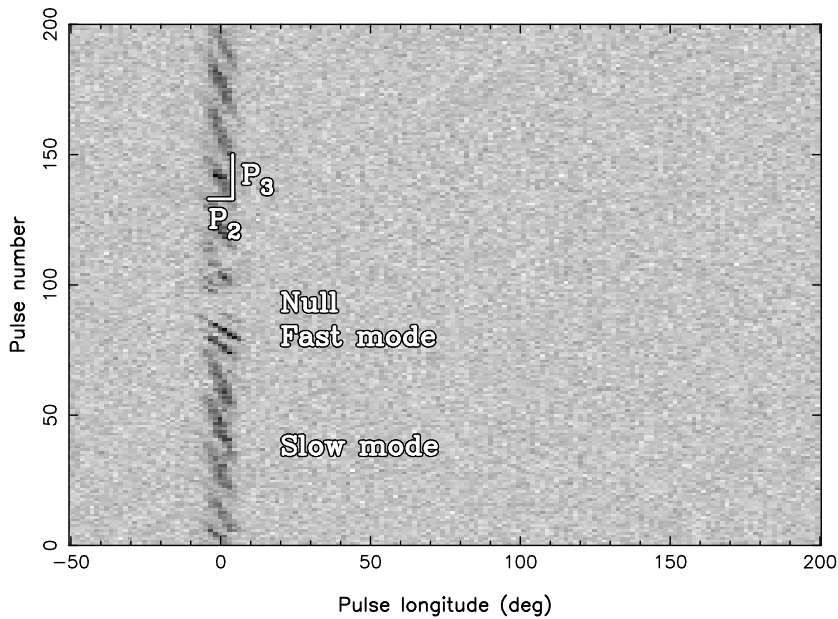}
}
\caption{A pulse-stack of successive radio pulses of PSR
  B1819$-$22. The radio intensity (darker is brighter) is plotted as
  function of rotational phase (or longitude) versus pulse
  numbers. The emission (in the subpulses separated by $P_2$) appear
  to ``march'' in rotational phase from pulse to pulse creating a
  distinct pattern which repeats itself every $P_3$ pulses, a
  phenomenon called ``drifting subpulses''. In addition this pulsar
  shows ``mode changes'', which in this case manifests itself as two
  distinct drift modes: the slow and fast mode. The pulsar also has a
  ``null'' state, i.e. for a few stellar rotations there was no radio
  emission emitted by the pulsar. This figure shows 1380 MHz data
  obtained by the Westerbork Synthesis Radio Telescope \citep{wes06}.}
\label{fig_stack}
\end{figure*} 
\subsection{Short timescale phenomena}
Radio pulsars display a myriad of amplitude modulation
effects, as seen in Fig. \ref{fig_stack}. Averaged over many rotations, most pulsars have a
reproducible pulse shape, reflective of the long-term stability of
their rotation and magnetic fields. However, pulse shape and intensity
can vary considerably in sequential rotations of a pulsar. Ordered and
stochastic processes affect some pulsars to varying degrees; ordered
modulations include sub-pulse drift (gradual phase shift of pulse
peak), mode changing (regular changes between two or three distinct
pulse shapes), and nulling (cessation of the radio beam). Pulse
statistics using full polarimetry \citep{scr+84,scw+84,kkj+02,mrg07}
have led to gradual improvement of our understanding of orthogonal
mode emission \citep{mck03b,es04}.  Drifting sub-pulses in PSR~B0943+10
led \cite{dr01} to champion the `rotating carousel' model for the
circulation of the pulsar beamlets \citep[see e.g.][]{rrvs06,mr08}.
However, it is far from clear if this model can be applied in every case
of drifting subpulses and other models have been proposed
\citep[e.g.][]{cr04a}.

It is also of particular interest to understand contributions of
pulse-to-pulse variability to precision pulsar timing, referred to as
jitter noise.  Correcting, or at least accounting for, jitter noise is
imperative for optimizing strategies for SKA pulsar timing activities
\citep{sod+14}.  The brightest MSPs at decimetre wavelengths,
PSR~J0437$-$4715 and J1713+0747, have been the most extensively
studied \citep{sc12,ovb+14}. While the emission of these two
millisecond pulsars show many similarities to that of slower pulsars,
many of the classic single pulse phenomena, such as drifting subpulses
or pulse nulling are not observed \citep{vam98}.

Pulse nulling, where the radio emission appears to switch off, was one
of the first phenomena identified \citep{bac70}, and is the
prototypical example of pulsar variability (see review by
\citealt{big92a}).  In the ``classical'' nullers, emission is observed to
switch off and on rapidly (few tens of pulses) and some pulsars had a
relatively high ``off'' fraction \citep{wmj07}.  It came as a major
surprise when \cite{mll+06} discovered pulsars, which only produce one
pulse of emission every tens of minutes (the so-called RRATs).
Several tens of these objects have now been found \citep{mlk+09}.
Because their duty cycle is so low, their population could rival that
of the normal pulsars \citep{kk08}.  

Kerr et al. (2014) show that several distinct timescales are present
in the PSR~J1717$-$4054, which has both long (many hours) and short
(few seconds) nulling periods.  The broadband nature of pulse nulling
was investigated by \cite{bgk+07}.  Our knowledge of how these various
populations fit together is given by \cite{bjb+12}.

Other effects such as intense giant pulses \citep{sr68,ccl+69} or
``giant micropulses'' \citep{jvkb01} occur in some pulsars at a
limited phase range. The energy distribution of radio pulses can
provide a window into the state of pulsar plasma and the method of
emission generation. There exist a great number of viable plasma-state
models, a few of which predict pulse energy distributions; the most
commonly-proposed predictions are of Gaussian, log-normal, and
power-law distributions. \cite{cjd01,cjd03a} and references therein,
provide discussion on these models. Energy distributions have now been
examined in detail for a number of pulsars \citep{bjb+12},
demonstrating that most pulsars obey log-normal statistics. These
analyses have substantially contributed to the hypothesis that genuine
``giant pulses'' are generated separately from standard pulse
generation and are much more rare. Giant pulses appear to have
power-law energy distributions, distinct from the otherwise log-normal
main pulse components \citep{lcu+95,jr02}.

It was giant pulses that led to the discovery of the Crab pulsar
\citep{sr68}.  Classical Crab pulsar giant pulses are characterized by
their appearance at the same phase range as the high-energy emission
\citep{lcu+95}, high flux densities \citep{hkw+03,ps07}, and power-law
intensity distributions \citep{ag72,lcu+95}, Since the discovery of the
Crab pulsar, other giant pulse emitters have been observed, both in
young pulsars (like for instance the Crab twin PSR~B0540-69
\citealt{jr03}), and in millisecond pulsars \citep{cst+96,rj01,kbm+06}.
Common aspects between all giant pulse emitting pulsars are currently
not known.  Giant pulses in the Crab seem to share properties such as
their short pulse widths down to 0.4~ns \citep{hkw+03,he07}, and high
brightness temperatures up to 10$^{39}$ K \citep{spb+04}, indicating
coherent emission mechanisms.
In addition to the giant pulses, bright, rare, `spiky' emission is seen in
other pulsars, for example the bursty emission in PSR~B0656+14 \citep{wsrw06},
the giant-like pulses from PSRs~J1048--5832 and J1709--4429 \citep{jr02}
and the giant micro-pulses from the Vela pulsar \citep{jvkb01}.

A number of tests of pulsar models can be performed using the
statistics of per-rotation pulsar modulation. \cite{wab+86} first
noted differences in modulation between core and conal-type pulse
profiles, while \cite{jg03} derived theoretical predictions for
anti-correlations between pulse-to-pulse modulation and four
``complexity parameters,'' corresponding to four pulsar emission
models. Their complexity parameters are: $a_1 = 5\dot
P^{2/6}P^{-9/14}$, for the sparking gap model, $a_2 = (\dot
P/P^3)^{0.5}$ for the continuous current outflow instabilities, $a_3 =
(P\dot P)^{0.5}$ for surface magnetohydrodynamic wave instabilities,
and $a_4 = (\dot P/P^5)^{0.5}$ for outer magnetospheric
instabilities. Few observational studies have been done to test these
effects, however the \cite{jg03} study for a small sample of core-type
profiles disfavoured the magnetohydrodynamic wave instability model,
and the study of $\sim$190 pulsars \citep{wes06,wse07} at 21 and 92\,cm
indicated that the modulation index is generally higher at lower
frequencies, and noted a weak correlation between modulation index and
age that is dampened at higher frequency.

The study of single-pulse modulation in a large pulsar sample can also
contribute to several practical questions, for instance: how common is
giant-pulse emission? Are the prospects of pulsar detection in other
galaxies better for single-pulse or Fourier searches? Quantification
of pulsars' modulation will also aid in understanding the physical
makeup of ``rotating radio transients'' \citep{mll+06}, which mostly
appear to represent an extreme case of nulling pulsar and may
contribute a problematically large contribution to Galactic pulsar
populations \citep{kk08}.

\subsection{Long timescale phenomena and pulsar timing}

As mentioned above, since the early days of pulsar astronomy, pulsars
have been known to change between modes on short (minute to hour) time
scales \citep{bac70b} and been extensively studied since.  Of
particular interest recently is the discovery that PSR~B0943+10
changes modes simultaneously in radio and in X-rays \citep{hhk+13}. The
X-ray state change from a thermal to a non-thermal component is
challenging for models of pulsar emission.

The question of pulse profile variation on the very long timescales
remains active. \cite{klo+06} identified the first of the so-called
intermittent pulsars, which was found to quasiperiodically change
between a 10 day radio-bright state and 30-day radio faint state.  The
timing data from PSR B1931+24 are well fit by a model with two rates
of spin-down, one for each emission phase. During the inactive phase,
the spin-down rate is $-10.8 \times 10^{-15}$ Hz s$^{-1}$, but
switches to $-16.3 \times 10^{-15}$ Hz s$^{-1}$ when active, an
increase of around 50\%.  Longer periods of quiescence have been
detected in PSRs~J1841--0500 and J1832+0029 with similar behaviour in
the timing of the pulsars \citep{crc+12}.

A less extreme version of correlated emission and rotation changes is
seen in a small number of pulsars identified by \cite{lhk+10}.  Six
pulsars are seen to show distinct emission states, identified by
long-term pulse-shape changes which are accompanied by correlated
changes in spin-down rate. The magnitude of the spin-down rate changes
produced is much smaller than is seen in the intermittent pulsars;
less than 10\% in each case.

The relationship between emission state and rotation rate is explained
by changing charged particle currents in the pulsar magnetosphere. A
global failure of these currents is thought to be responsible for the
inactive phase, creating a dearth of charged particles at the magnetic
poles. When the outflowing particles are present and producing radio
emission, they also provide an additional torque, resulting in the
increased spin-down rate seen during the active phase.

Glitches appear to have at least some effect on the pulse profile
(for the latest papers on observed glitches see \citealt{elsk11,ymh+13}).
For PSR~J0742--2822, \cite{ksj13} found
that the degree of correlation was influenced by a glitch occurring
midway through the dataset; a strong correlation was found after the
glitch, with no correlation before.  Glitches have previously been
linked to emission changes in radio pulsars and in magnetars
\citep{ccr+07,wje11}.

A further example of correlated changes in emission and rotation is
seen in PSR~J0738--4042 \citep{bkb+14}.  In 2005, the pulsar exhibited
a sudden change in pulse-shape, which had been relatively stable for
at least 16 years prior, and occurred simultaneously with a 15\% drop
in the pulsar's spin-down rate. The changes were attributed to
variations in magnetospheric current flow, hypothesised to have been
due to an influx of external material, such as an asteroid
\citep{bkb+14,scm+13}. \cite{lgw+13} report on a 20 year dataset of the
Crab pulsar which appears to show that the magnetic axis is slowly
migrating away from the axis of rotation in contrast to population
studies \citep{tm98,wj08a} which conclude the opposite.

What is the link between the nanosecond emission, the microstructure
and the components in a pulse profile? Why do some pulsars produce
giant pulses, while others null?  What sets the on and off null
durations? How do pulsar profiles change with age and do glitches play
a role in this? These questions are fundamental to our understanding
of the emission mechanism, and should be addressed by high cadence
monitoring of many hundreds of sources in the SKA era. Furthermore,
the entire question of pulse profile stability, one of the shibboleths
in pulsar astronomy, has been called into question over the past
decade. There now appears to be a clear link between rotational
instabilities and profile changes, though questions of cause and
effects remain unclear. In addition it seems possible that some
changes are externally driven while some remain intrinsic to the pulsar
itself. Pulse profile stability is crucial to high precision timing
and the quest for gravitational waves and therefore understanding
profile stability is a key priority for the SKA.

\section{Conclusions and early science}
The SKA brings high sensitivity over a broad frequency range, ideal for
the study of broadband, highly variable, weak sources such as pulsars.
In the SKA era, great strides will be made in understanding the location
and behaviour of the coherent radio emission from pulsars. In turn, this
understanding of pulsar emission will be crucial for the detection and
study of gravitational waves, surely to be one of the crowning glories
of SKA science.

As demonstrated above, the science enabled by SKA studies of pulsar
emission and magnetospheres is impressive.  For the first time since
the discovery of pulsars, we have the chance to find the ``Rosetta
stones'' of pulsar research that will allow us to infer the
fundamental emission properties and their origin as well as the
conditions in the magnetosphere. The importance of just a few sources
for our understanding has been demonstrated, e.g., by the few known
intermittent pulsars and the double-pulsar. The SKA will deliver more
of those and other types of radio emitting neutron stars
\citep{tkb+14, kbk+14}. At the same time, the sheer number of pulsars
that the SKA will be able to study in exquisite detail promises to
make real advances in solving the decade-old ``pulsar problem''. This
will be aided by observations with other facilities at other
wavelengths. 

The key in advancing the radio studies will however be the supreme
sensitivity of the SKA. Clearly, already SKA1 (LOW and MID) will be
superb tools. Eventually, more sensitivity with full SKA will help
even more. In terms of early science with SKA1, this hinges on early
new pulsar discoveries and the telecsope sensitivity during the early
stages. It is difficult to estimate precisely what can be achieved
during this phase, as the impact will be different for different
sources. Nevertheless, the science proposed here can be addressed as
long as the telescope has the characteristics of bandwidth,
polarization purity and sensitivity required to study in detail the
sources discovered. It is, however, the full SKA-1 that will really
enable substantial progress in this area, given the large number of
new sources including pulsars that can be studied in detail with full
polarization, in single pulses and at regular intervals to understand
variability. Given that the SKA will probably be the largest radio
telescope built for a while, we should not miss this golden
opportunity to solve one of the great mysteries in astrophysics.

\section*{Acknowledgements}
J-PM acknowledges work supported through the Australian Research Council grant DP140104114.

\bibliographystyle{apj}
\bibliography{journals_apj,modrefs,psrrefs,crossrefs} 
\end{document}